# Three-Dimensional Automated Assessment of the Distal Radioulnar Joint Morphology according to Sigmoid Notch Surface Orientation

1. Roner, Simon, MD [1, 2]
2. Fürnstahl, Philipp, PhD [1]
3. Scheibler, Anne-Gita, MD [2]
4. Sutter, Reto, MD [3]
5. Nagy, Ladislav, MD [2]
6. Carrillo, Fabio, PhD [1]

*(1) Research in Orthopedic Computer Science, Balgrist University Hospital, University of Zurich, Zurich, Switzerland*
*(2) Department of Orthopaedics, Balgrist University Hospital, University of Zurich, Zurich, Switzerland*
*(3) Department of Radiology, Balgrist University Hospital, University of Zurich, Zurich, Switzerland*

**Corresponding author:**
Dr. Sc. Fabio Carrillo
Fabio.Carrillo@balgrist.ch
Research in Orthopedic Computer Science
Balgrist University Hospital
University of Zurich
Switzerland

## Abstract

**Purpose**

The aim of this study was to develop a new method for generating reproducible 3D measurements for the quantification of the distal radioulnar joint morphology. We hypothesized that automated 3D measurement of the ulnar variance and the sigmoid notch angle are comparable to those of the gold standard, while overcoming some of the drawbacks of conventional 2D measurements.

**Methods**

Radiological data of healthy forearm bones (radiographs and CT) of 53 adult subjects were included in the study. Automated measurements for the assessment of the sigmoid-notch morphology based on 3D landmarks were developed incorporating the subject-specific estimation of the cartilage surface orientation. A common anatomical reference was defined among the different imaging modalities and a comparison between the sigmoid notch angle and UV measurements was performed in radiographs, CT scans and 3D models. Finally, the developed UV measurements in 3D were compared to the method by radiographs in an experimental setup with 3D printed bone models.

**Results**

The proposed automated 3D analysis of notch subtype showed a significantly larger (p = 0.04) notch radius for negative notch angle (mean 22.4 mm +/-14.4) compared to positive sigmoid notch angle subjects (mean 16.9 mm +/-5.1). Similar UV measurements were obtained in healthy joint morphologies with a high correlation between the radiographs and 3D measurements, for sigmoid notch angle (0.77) and UV (0.85). In the experimental setup with a modified radial inclination, the UV was on average 1.13 mm larger in the radiographs compared to the 3D measurements, and 1.30 mm larger in the cases with a modified palmar tilt. Furthermore, UV radiograph measurements on the modified palmar tilt deviated from the expected linear behavior observed in the 3D measurements.

**Conclusion**

The developed 3D measurements allowed to reliably quantify differences in the sigmoid notch subtypes. Since the method is not dependent on the position and orientation of the distal radius, the hereby provided groundwork can be used for the accurate quantification of distal radioulnar joint pathologies.

**Introduction**

In the area of hand surgery, computer-assisted (CA) three-dimensional (3D) analysis of bone anatomy has become a valuable tool for patient-specific preoperative planning of orthopedic surgeries, allowing the exact quantifications of the malunion or deformation[1-3], and the precise surgical navigation by means of patient-specific instruments (PSI). For surgical procedures involving the length correction of the distal radioulnar joint (DRUJ), e.g., fracture reconstruction, extra and intra-articular corrective osteotomies, the accurate measurement of the joint morphology is needed in order to avoid DRUJ impingement, joint instability and other associated complications [4-7]. Two of the common measurements for the radiological assessment of the DRUJ morphology are the ulnar variance (UV) and the sigmoid notch angle (SNA). The UV is defined as the length difference between the most distal part of the ulnar head and the most proximal part of the radiocarpal joint, measured on postero-anterior plain radiograph[8]. The SNA can be measured on the coronal plane as the angle between the longitudinal shaft of the ulna and the tangent to the sigmoid notch of the radius, as previously described by Ekenstam[9] and Sagerman[10].

Additionally, the sigmoid notch type of the radius is commonly classified by Förstner[5] in three orientations according to the SNA (**Figure 1**); positive angles correspond to conical shapes (type 1), neutral angles to cylindrical shapes (type 2) and negative angles to hemispherical shapes (type 3). A correlation between SNA and UV was propagated for the healthy anatomies, namely a positive SNA correlates with a negative UV, neutral SNA with neutral UV, and negative SNA with positive UV[5]. Several authors have demonstrated this correlation through the evaluation of conventional two-dimensional (2D) radiographs and 2D cutting surfaces in cadaveric studies[4,6].

Once 3D data became available through computed tomography (CT) and magnetic resonance imaging (MRI), new UV methods were developed[11]. Morphological analysis was extended to the axial plane by measuring the radius of the sigmoid notch (SN radius) and the ulnar head and validating those parameters through cadaver studies[6,11]. The availability of 3D morphological data of the DRUJ also enabled the 3D preoperative planning of corrective osteotomy, fracture reconstruction, arthroplasty, tumor resection and other complex forearm procedures[12-16]. Nevertheless, the currently available DRUJ measurements methods (i.e., UV, SNA and SN radius measurements) are still 2D-based, and the complexity of the DRUJ morphology can be often oversimplified, hindering the benefits of the 3D planning.

For the optimization of 3D planning of the forearm, 3D DRUJ morphological measurements methods are needed that are able to accurately describe the joint morphology and to account for deformities which might be only evident in the 3D space[17-19]. Only few works have aimed to measure the DRUJ morphology in 3D, by measuring the UV in bone models obtained through segmentation of CT images [19-21]. Although these methods are promising, there are still some challenges to be solved for their clinical applicability. A summary of the related work and its associated challenges are given in **Table 1.**

In this study, we develop new methods for the quantification of the DRUJ morphology using inherent and automated 3D methods, which are hitherto still missing for the definition of a surgical 3D planning of the DRUJ [22], and required to overcome the drawbacks of the aforementioned 2D measurements. Thereby, we describe the new 3D methodology and we perform a validation against existing gold standards DRUJ quantification methods.

## Materials and Methods

### Demographics

A retrospective analysis was performed on patients treated at our hand surgery department, which had CT images of their healthy wrist for comparison and 3D planning. Inclusion criteria were the availability of radiological data of the contralateral healthy forearm comprising plain radiographs and CT data. In total, 58 subjects being 18 years or older were included. One subject was excluded due to noticeable deformities of the distal radius, and four patients were excluded because of insufficient image quality of the CT images. The measurements were performed on the remaining 53 subjects. The subjects had a mean age of 34.0 years (18.6 – 66.4) and a gender distribution of 35 male (66%) and 18 female (34%) subjects. In total, 23 left (43.4 %) and 30 (56.6 %) right pairs of forearm bones were analyzed.

### Data description

At our department, standard postero-anterior radiographs of the wrist were taken in neutral position, with the elbow elevated to shoulder level and flexed to 90°, with a standard digital radiography system (Ysio; Siemens Healthineers, Erlangen, Germany) and an isotropic image resolution of 139 μm). The center of the field of view was aligned to the radiocarpal joint center. The CT images of the forearm were obtained in neutral forearm rotation in superman position (slice thickness 1 mm; 120 kV; Philips Brilliance 40 CT; Philips Healthcare, Eindhoven, The Netherlands). To allow comparison between radiographs, CT image and 3D bone model measurements, an anatomical axis on the distal third of the radius was defined and used for the realignment of the CT data, i.e., the CT axes were realigned to match an anatomical orientation in correspondence to the plane of an AP radiograph. The 3D triangular surface models of the bones were generated by segmenting the bones from CT data in a semi-automatic fashion as described by Fürnstahl et al.[12] and in accordance to the clinical standards used in our institution for the generation of patient-specific surgical navigation instruments.

**Measurements**

The SNA was measured on plain radiographs according to the method described by Förstner[5], described in **Figure 2A**. Similarly, the UV was measured using the method of perpendiculars[23], which is an advancement of the technique described by Palmer[8,24], shown in **Figure 2B**. Positive UV-values represent an ulnar head point being located more distal than the sigmoid notch center of the radioulnar corner.

On coronal CT reconstructions, the SNA was measured as the angle between the long bone axis of the distal radius and a line defined along the sigmoid notch, being a positive angle towards ulnar and negative towards radial side, respectively. To measure the UV, a line perpendicular to the baseline was drawn through the edges of the center of the sigmoid notch. A second parallel line was drawn to define the second UV measurement point, which crossed the most distal part of the ulnar head. The distance between the two lines defined the UV on CT reconstructions.

The 3D bone models were imported to the in-house preoperative planning software to perform the measurements according to the developed 3D measurement method. Although the in-house planning software does facilitate the measurement process through dedicated planning tools, the same measurements could be done using any dedicated CAD software. Afterwards, an anatomical and right-handed coordinate system was aligned on the distal third of the radius and the ulna pointing proximally in the x-axis, as illustrated in detail in **Figure 3**. The y-axis of the radius coordinate system was aligned between the radial styloid and the center of the sigmoid notch, representing the radio-ulnar direction. The z-axis defined the palmar-dorsal axis and represents the xy-plane, in which the measurement of the SNA is performed, i.e. SNA is measured with respect to the rotation along the palmar-dorsal (z).axis (**Figure 3B**).

In our 3D measurement method, the sigmoid notch surface was described by anatomic landmarks surrounding the estimated cartilage area, which were manually defined on the subchondral bone surface of the sigmoid notch of the radius (SN Landmarks, **Figure 4A**). Subsequently, a two-step fitting algorithm was applied to find the best fitting cylinder to the SN Landmarks. In the first step, a geometric-based fitting algorithm[25] was applied to automatically fit a sphere to the given 3D landmark points. In the second step, a least-square fitting (LSF) algorithm, based on Gauss-Newton optimization[26,27], is used to fit a cylinder to landmarks of the sigmoid notch. The LSF algorithm is initialized using the radius of the previously fitted sphere, and the x-axis of the defined coordinate system of the distal ulna (red axis on the ulna, **Figure 3A**), as a first approximation of the radius and the long axis of the cylinder, respectively. The output of the algorithm was a cylinder fitted to the described landmarks (SN-fitted cylinder in cyan, **Figure 4B-C**). The SNA was measured as the angle between the main axis of SN-fitted cylinder (black axis, **Figure 4D**) and the long-bone axis of the radius (red axis, **Figure 4D**), projected on the anterior-posterior plane of the anatomic coordinate system defined by the z-axis (xy-plane) in **Figure 3B**.

The 3D UV is measured as the distance between the most proximal point of the sigmoid notch edge in the radiocarpal joint along the axial plane of the radius (magenta sphere, **Figure 5**), and the most distal point of the ulnar head along the axial plane of the ulna (green sphere, **Figure 5**), with respect to the x-axis of the radius.

**Experimental setup**

To investigate the influence of DRUJ incongruence on the existing measurement methods, we executed an experimental setup with 3D printed bone models of the distal radius and ulna (**Figure 6**). The radial inclination and the palmar tilt from a healthy bone model were modified according to the previously described anatomic coordinate system of the distal radius. The goal of the setup was to simulate the influence of a pathological radial inclination and palmar tilt on the measurement methods in radiographs and in 3D. Based on a healthy forearm bone model, the radial inclination was modified by rotating in both directions in steps of 10° from -10° up to 50° in the anatomic coordinate system, with a healthy radial inclination defined at 20°. Similarly, the palmar tilt was modified from -20° up to +40°, with a healthy palmar tilt defined at 10°. The UV measurements in all simulated positions were performed in a standardized setup (**Figure 6B**) and by two independent readers on radiographs (**Figure 6C-D**) of 3D printed bone models (**Figure 6A**). Additionally, the measurements were performed automatically on the 3D bone models as previously described.

**Statistical Evaluation**

The correlation coefficient was calculated by the Pearson method. Descriptive statistics were used to analyze differences between the measurement methods. After graphical testing for normal distribution, a paired t-test was performed for analysis of the difference between UV and SNA in different image modalities; the level of significance was established at $p < 0.05$.

**Source of Funding**

This work was partially funded by the Swiss national science foundation (SNF) project No. 325230L_163308.

**Results**

The morphology analysis of the sigmoid notch showed an average SN radius of 15.0 mm (+/- 3.7) in the CT method, and 16.9 mm (+/- 5.1) in the 3D method for cases with positive SNA, as illustrated in **Figure 7**. In the case of a negative SNA, the mean SN radius was 17.5 mm (+/- 4.7) in the CT method, and 18.9 mm (+/- 5.8) in 3D method. A significant larger SN radius was measured in subjects with a negative SNA in comparison to subjects with a positive SNA in both measurement methods, the 3D ($p = 0.04$) and the CT ($p = 0.03$). The 3D measurements of the SN radius were significantly larger ($p = 0.03$) compared to the CT measurements of the same subjects. One outlier measurement was detected in the 3D method with a radius of 70.3 mm. This case was verified, and the large radius was confirmed to be a consequence of the specific DRUJ anatomy.

In **Table 2** we provide an overview of the individual Spearman correlation of both measurements, SNA and UV, among the different imaging modalities. An additional, correlation analysis between the measured values of the SNA and UV was performed among the different imaging modalities, showing a moderate negative Spearman correlation in both, the radiographs method (-0.62, $p < 0.01$), and the 3D method (-0.61, $p < 0.01$). A stronger negative correlation between SNA and UV values was observed in the measurements of coronal CT images (-0.69, $p < 0.01$).

Additionally, measurements from the experimental setup on the 3D models with a modified radial inclination showed a larger UV for the radiograph method (**Figure 8A**), in average 1.13 mm, with respect to the 3D measurements. Values of the UV had an expected decreasing behavior from the negatives (-10°) to the positive angles (+50°) of the radial inclinations of -0.25 mm/° for the 3D measurements, and -0.24 mm/° for the radiograph method. In the case of a modified palmar tilt (**Figure 8B**), the radiographs UV measurements were in average 1.30 mm larger than those of the 3D methods. Moreover, an expected increase of the UV measurements was observed with respect to the palmar tilt angles (+0.04 mm/° in 3D and

+0.02 mm/° in radiographs). However, in the case of the radiograph measurements, values of the UV measurements did not exhibit a linear behavior.

## Discussion

The goal of this study was to investigate new methods for the 3D measurement of the DRUJ morphology with the aim of improving the available tools for the 3D planning of forearm orthopedic surgeries. We have developed automated 3D methods for the quantification of the SNA and the UV of the DRUJ, which are more observer-independent than the current gold standard methods.

We were able to quantify for the first time the difference in the sigmoid notch subtypes with respect to the DRUJ orientation propagated by Förstner [5] (**Figure 1**) and further investigated by Tolat et al [6], confirming the hypothesis of morphology differences that were not possible to be demonstrated previously through radiographic measurements. Specifically, our results demonstrated size differences of SN radius among the different sigmoid notch subtypes with respect to the corresponding SNA. Namely, a significant larger SN radius was measured in both, the 3D and the CT measurement methods for negative SNA, compared to the SN radius of positive SNA subjects. Furthermore, we were able to confirm in the 3D bone models the negative correlation between UV and SNA previously described in radiographs and coronal CT images [5,10]. This moderate correlation between the UV and the SNA in the long bone axis was an argument by several authors outlining the pitfalls of considering the sigmoid notch orientation in modifications of the ulnar length to avoid DRUJ impingement [4,5,10]. Consequently, incorrect or various measurements of UV can easily limit the planning and execution of corrective osteotomies of the forearm bones by an inaccurate quantification of the DRUJ morphology through 2D-based methods.

To the best of our knowledge, only few descriptions of UV measurements exist on 3D bone models. In 2014, Kawanishi performed a study on the comparison of 2D radiographs versus 3D CT-measurements during forearm rotation [19]. The UV was measured by the plane perpendicular to the longitudinal axis of the entire radius as the distance between the plane passing through the most distal point of the ulnar head and the plane passing midway of the lunate fossa. Fu et al, demonstrated a different method by measuring the UV as the distance along the ulnar z-axis between the center of the radial coordinate system and the center of the ulnar coordinate system, both as long bone axis[21]. Both methods use the long bone axis as measurement reference; therefore, comparison with the commonly used radiograph method is not applicable. In 2016, Daneshvar et al. described anatomic oriented measurements for the quantification of the DRUJ morphology and a correlation analysis between the UV and SNA[20]. However, the measurements were applied only two-dimensionally on the corresponding sequences, e.g. quantifying the SNA on the mid-coronal plane by manual measurements. Our 3D measurements of UV and SNA were performed in relation to the 3D distal axis of the radius, which have the advantage of being directly comparable to the gold standard method.

In our experimental setup, UV measurements were obtained in radiographs and on 3D bone models with simulation of different radial inclinations and palmar tilts. The observed linear decreasing behavior of the UV in relation to the radial inclination can be explained geometrically, as the radial inclination directly modifies the measurement points of the UV. However, due to a rotated and tilted center of the sigmoid notch edge through the corner point between the maximally radiolucent lines of the ulnar fossa and the sigmoid notch, the measurements of the UV on radiographs are difficult and can deviate from the position-independent measurements of 3D bone models, as observed in our study with a de-rotated palmar tilt (**Figure 8B**). Finally, the developed automated 3D methods can already partially solve the inter-rater variability reported in 2012 by Laino[11] with a large average bias from manual UV measurements in different image modalities.

The ultimate clinical goal of DRUJ morphology measurements is the accurate approximation of the bone cartilage area. In our opinion, dedicated 3D methods for DRUJ morphology analysis can help solve the known drawbacks and inaccuracies of manual 2D measurements and can better approximate the cartilage area, enabling an improved diagnosis and treatment for patients. Although the herein developed 3D measurement can already solve some of the limitations of the existing 2D-based measurement techniques, our study had some limitations. First, our study is based on retrospective clinical data. For the clinical validation and standardization of the developed 3D measurements, a corresponding prospective study with a control group is needed. Second, the estimated cartilage area is described by means of subchondral-bone landmarks placed on the perimeter of the sigmoid notch. These landmarks are not able to include all the information provided by the entire sigmoid notch surface and could therefore affect the sensitivity of the method. Also, the landmarks used for the calculation of DRUJ measurements were manually placed by an expert orthopedic surgeon, and are, therefore, dependent on the surgeon criteria. One possible solution to these limitations is the use of automated techniques for region and landmark identification such as statistical shape models $^{28-31}$ or curvature feature identification$^{32,33}$.

In conclusion, the developed 3D methods are not dependent on the position or orientation of the distal radius and can thereby be used for the computer-assisted planning and treatment of complex DRUJ pathologies. From our experience, patients that benefit the most from computer –navigated surgeries are those with multi-planar deformities, which are not possible to accurately assess by standard 2D-techniques. In our group, we have systematically translated standard anatomical measurements into the 3D space, and therefore the translation of the DRUJ measurements is part of this effort to improve the 3D preoperative planning process. Future studies should further investigate the optimal congruency of the sigmoid notch, the ulnar head, and the ulnocarpal joint by more accurate analyses in 3D bone models. In our opinion, future efforts should focus on obtaining realistic accurate 3D models of the cartilage area around the DRUJ. This could be done by automatic cartilage segmentation through learning-based methods.

**Figures legends**

**Figure 1**

The distal radioulnar joint surface orientation propagated by Förstner in subjects with cylinder orientation in neutral UV (**A**), with a spherical joint surface in positive UV subjects (**B**) and in negative UV subjects with inclined cylinder orientation (**C**).

**Figure 2**

SNA measurement method for radiographs.

(**A**) Measurement of SNA between the long bone axis (line 1) of the distal radius and the sigmoid notch orientation line (line 2). The orientation of the sigmoid notch is positive when line 2 is oriented towards the ulna head (as presented), or negative when oriented towards radial, with respect to line 1, similar to the description made by Sagerman[10].

(**B**) UV is measured as the distance between lines 3 and 4. Line 3 is the radial reference line, orthogonal to line 1, and which crosses the center of the sigmoid notch edge through the corner point between the maximally radiolucent lines of the ulnar fossa and the sigmoid notch of the radius. Line 4 represents the ulnar reference line, which is parallel to line 3, and drawn over the most distal point of the ulnar head. The UV is negative if line 4 is proximal to line 3, (as depicted), or positive otherwise.

**Figure 3**

Definition of coordinate system for the DRUJ measurements in 3D

(**A**) 3D anterior-posterior view and (**B**) axial view of the distal third of the radius and ulna and their corresponding coordinate system. Red arrow indicates the longitudinal direction or x-axis, green arrow the radioulnar direction or y-axis and the blue arrow indicates the palmar-dorsal direction or z-axis, on each bone respectively.

**Figure 4**

Definition of the sigmoid notch angle (SNA) in 3D.

**(A)** Sigmoid notch surface described by anatomic landmarks (SN landmarks, blue spheres) on the subchondral bone.

**(B)** Example of the resulting cylinder fitted to the SN Landmarks (SN-fitted Cylinder in Cyan) and its corresponding main axis (shown in black).

**(C)** Coronal view of the SN-fitted cylinder

**(D)** The SNA was measured as the angle between the main axis of the SN-fitted cylinder (black axis) and the x-axis of the radius (red axis), projected to the anterio-posterior plane of the anatomic coordinate system defined in **Figure 3A**.

**Figure 5**

Definition of the ulnar variance (UV) measurement in 3D.

UV is measured along the long-bone axis of the radius (x-axis in red) as the distance between the most proximal point of the sigmoid notch edge of the radius in the radiocarpal joint (magenta sphere) and the most distal point of the ulnar head along the long-bone axis of the ulna (green sphere).

**Figure 6**

Experimental setup.

**(A)** 3D-printed bone models with modified palmar tilt from -20° to +40. 3D-printed bone model with healthy palmar tilt is indicated (T 10°)

**(B)** 3D-printed bone models with modified radial inclination from -10° to +50. 3D-printed bone model with healthy radial inclination is indicated (I 20°)

**(C)** Measurement setup with 3D-printed distal forearm bone models. Model shown correspond to the healthy model.

**(D)** Posteroanterior radiographs of healthy 3D-printed distal forearm bone model with ulna variance measurement (shown by the yellow lines).

**(E)** Posteroanterior radiographs of malunited 3D-printed distal forearm bone models at the exact same position compared to the healthy bone model in (C).

**Figure 7**

Boxplot of mean sigmoid notch radius (SN Radius) on positive (POS) and negative (NEG) SNA measured by the 3D and the CT method.

**Figure 8**

Comparison between 3D (bold line) and radiograph measurements (dotted line) of the ulna variance with respect to the modification of angles in the printed 3D models of the DRUJ corresponding to the radial inclination **(A)** and the palmar tilt **(B)**. Values of the healthy radial inclination ($I_h$) and palmar tilt ($T_h$) are indicated in bold, and UV measurements are shown by a triangle for the 3D method, and a square for the radiographs.

**Table 1.-** Overview of current available methods for quantification of DRUJ according to image modality.

| Image Modality | Measurement technique | Sigmoid notch (SN) measurements | Drawbacks | Related Articles |
|---|---|---|---|---|
| **Radiographs** | Manual performed 2D geometry (lines, circles,..) | **UV, SNA** Coronal plane along distal bone axis **SN radius** Not possible | **Acquisition:** <br>• Projection of anatomy <br>• Manual setup <br>**Measurement:** <br>• Manually performed <br>• Non reproducible (observer dependent) | Förstner[5] <br> Tolat[6] <br> De Smet[4] |
| **Multiplanar sliced data CT / MRI** | Manual performed 2D geometry (lines, circles,..) | **UV, SNA** Coronal slice along distal bone axis **SN radius** Axial slice | **Acquisition:** <br>• Manual (anatomical) axis alignment <br>**Measurement:** <br>• Manually performed <br>• Non reproducible (observer dependent) | Laino[11] |
| **3D surface mesh bone models** | Manual aligned 3D geometry projected on one plane | **UV** Projected coronal plane along long bone axis **SNA and SN radius** Not measured | **Acquisition:** <br>• Dependent on slice resolution <br>**Measurement:** <br>• Manually performed <br>• Not comparable with radiographic method | Kawanishi[19] <br> Fu[21] <br> Daneshvar[20] |

4  **Table 2.-** Correlation of sigmoid notch angle (SNA) and ulna variance (UV) measurement among different image modalities.

| Image Modality | SNA | UV |
|---|---|---|
| **Radiographs vs. CT** | 0.92 | 0.94 |
| **Radiographs vs. 3D** | 0.77 | 0.85 |
| **CT vs. 3D** | 0.84 | 0.85 |

6